\documentclass[12pt]{amsart}
\usepackage{graphicx,amsmath,amssymb}

\newcommand{\C}{\mathbb C}
\newcommand{\R}{\mathbb R}
\newcommand{\Z}{\mathbb Z}
\newcommand{\N}{\mathbb N}
\renewcommand{\d}{\prime} 
\newcommand{\dd}{{\prime \prime}}

\renewcommand{\Re}{{\rm Re}\,}
\renewcommand{\Im}{{\rm Im}\,}
\newtheorem{theorem}{Theorem}
\newtheorem{lemma}[theorem]{Lemma}
\newtheorem{proposition}[theorem]{Proposition} 
\newtheorem{corollary}[theorem]{Corollary}

\newtheorem{definition}{Definition}
\newtheorem*{remark}{Remark}
\newtheorem*{remarks}{Remarks}

\begin{document}
\title[]
{On the reality of the eigenvalues for a class of $\mathcal{PT}$-symmetric oscillators}
\author[]
{K. C. Shin}
\address{Department of Mathematics, University of Illinois, Urbana, IL 61801}
\date{January 20, 2002}

\begin{abstract}
We study the eigenvalue problem $-u^\dd(z)-[(iz)^m+P(iz)]u(z)=\lambda u(z)$ with the boundary conditions that $u(z)$ decays to zero as $z$ tends to infinity along the rays $\arg z=-\frac{\pi}{2}\pm \frac{2\pi}{m+2}$, where $P(z)=a_1 z^{m-1}+a_2 z^{m-2}+\cdots+a_{m-1} z$ is a real polynomial and $m\geq 2$. We prove that if for some $1\leq j\leq\frac{m}{2}$, we have $(j-k)a_k\geq 0$ for all $1\leq k\leq m-1$, then the eigenvalues are all positive real. We then sharpen this to a slightly larger class of polynomial potentials.
 
In particular, this implies that the eigenvalues are all positive real for the potentials $\alpha iz^3+\beta z^2+\gamma iz$ when $\alpha,\,\beta,\,\gamma \in \R$ with $\alpha\not=0$ and $\alpha \,\gamma \geq 0$, and with the boundary conditions that $u(z)$ decays to zero as $z$ tends to infinity along the positive and negative real axes. This verifies a conjecture of   Bessis and Zinn-Justin.
\end{abstract}

\maketitle

\begin{center}
{\it Preprint.}
\end{center}

\baselineskip = 18pt

\section{Introduction}
\label{introduction}
\subsection{The main results}
We are considering the eigenvalue problem
\begin{equation}\label{ptsym}
-u^\dd(z)-[(iz)^m+P(iz)]u(z)=\lambda u(z)
\end{equation}
with the boundary conditions that $u(z)$ decays to zero as $z$ tends to infinity along the rays $\arg z=-\frac{\pi}{2}\pm \frac{2\pi}{m+2}$,  where $m\geq 2$, $\lambda\in \C$ and $P$ is a real polynomial of the form
\begin{equation}
P(z)=a_1z^{m-1}+a_2z^{m-2}+\cdots +a_{m-1}z,\quad\text{with all}\quad a_k\in \R.
\end{equation}
The boundary conditions here are those considered by Bender and Boettcher \cite{Bender}. Note  that the boundary conditions for $m=3$ are equivalent to the conditions that $u$ decays to zero as $z$ tends to infinity along the positive and negative real axes. 
If a non-constant function $u$ along with a complex number $\lambda$ solves (\ref{ptsym}) with the boundary conditions, then we call $u$ {\it an eigenfunction} and $\lambda$ {\it an eigenvalue.} 

Before we state our main theorem, we first introduce some known facts by Sibuya \cite{Sibuya} about the eigenvalues $\lambda$ of (\ref{ptsym}), facts that hold even when $a_k\in \C$.
\begin{proposition}\label{main2}
The eigenvalues $\lambda_k$ of (\ref{ptsym}) have the following properties.
\begin{enumerate}
\item[(I)] Eigenvalues are discrete.
\item[(II)] All eigenvalues are simple.
\item[(III)] Infinitely many eigenvalues exist.
\item[(IV)] Eigenvalues have the following asymptotic expression
\begin{equation}\label{Bender_exp}
\lambda_k =\left(\frac{\Gamma\left(\frac{3}{2}+\frac{1}{m}\right)\sqrt{\pi}\left(k-\frac{1}{2}\right)}{\sin \frac{\pi}{m}\Gamma\left(1+\frac{1}{m}\right) } \right)^{\frac{2m}{m+2}}[1+o(1)]\quad\text{as $k$ tends to infinity},\quad k \in \N,
\end{equation}
where the error term $o(1)$ could be complex.
\end{enumerate} 
\end{proposition}
We will give precise references for Proposition \ref{main2} after Proposition \ref{prop} in Section \ref{properties}.
 In this paper, we will prove the following theorem that says that the equation (\ref{ptsym}) with a polynomial potential in a certain class has positive real eigenvalues only.
\begin{theorem}\label{main}
Let $a_k$'s be the coefficients of the real polynomial $P(z)=a_1z^{m-1}+a_2z^{m-2}+\cdots+a_{m-1}z$.
If for some $1\leq j\leq\frac{m}{2}$, we have $(j-k)a_k\geq 0$ for all $k$, then the eigenvalues of (\ref{ptsym}) are all positive real. 
\end{theorem}
\begin{corollary}\label{main3}
In particular, with $m=3$ the eigenvalues $\lambda$ of 
$$-u^\dd(z)+(iz^3+\beta z^2+\gamma i z)u(z)=\lambda u(z),\quad u(\pm\infty+0i)=0,$$
 are all positive real, provided $\beta\in \R$ and $\gamma\geq 0$.
\end{corollary}
\begin{proof}
This is a special case of Theorem \ref{main} with $m=3$, $j=1$ and $P(z)=\beta z^2-\gamma z$.
\end{proof}
We also mention that Delabaere et al. \cite{Pham,Delabaere} studied  the potential $iz^3+\gamma iz$ and showed that a pair of  
non-real eigenvalues develops for large negative $\gamma$. And  Handy et al. \cite{Handy2,Handy1} showed that the same potential admits a pair of non-real eigenvalues for small negative values of $\gamma\approx -3.0$.   
\begin{remark} {\rm By rescaling, the conclusion of Corollary \ref{main3} holds for the potential $\alpha iz^3+\beta z^2+\gamma iz$ when $\alpha \in \R-\{0\},\, \beta\in \R$ and $\alpha\, \gamma\geq 0$.}
\end{remark}

This paper is organized as follows. In the rest of Introduction, we will briefly mention some earlier work. Then in the next section, we state some known facts about the equation (\ref{ptsym}) and examine further properties. In Section \ref{main_proof}, we prove Theorem \ref{main}, and in Section \ref{extension_sec} we extend Theorem \ref{main}. Finally, in the last section  we discuss some open problems for further research.

\subsection{Motivation and earlier work}
Around 1995,  Bessis and Zinn-Justin conjectured that eigenvalues of 
\begin{equation}\label{zinn}
\left[-\frac{d^2}{dz^2}-\alpha(iz)^{3}+\beta z^2\right]u(z)=\lambda u(z),\quad\text{for}\quad \alpha\in \R-\{0\},\,\,\beta\in \R,
\end{equation}
are all positive real. And later Bender and Boettcher \cite{Bender} generalized the BZJ conjecture; that is, they argued that eigenvalues of 
\begin{equation}\label{Bender_problem}
\left[-\frac{d^2}{dz^2}-(iz)^{m}+\beta z^2\right]u(z)=\lambda u(z),\quad\text{for}\quad \beta \in \R,
\end{equation}
are all positive real when $\beta\geq 0$. Notice this follows for $\beta\leq 0$ by Theorem \ref{main} with $P(z)=\beta z^2$. The case $\beta>0$ is open, except for $m=3,\,4$ which are covered by Theorem \ref{main}.

Recently, Dorey et al.\ \cite{Dorey3,Dorey} have studied the following problem
\begin{equation}\label{bessis}
\left[-\frac{d^2}{dz^2}-(iz)^{2M}-\alpha(iz)^{M-1}+\frac{l(l+1)}{z^2}\right]u(z)=\lambda u(z),
\end{equation}
 with the boundary conditions same as those of (\ref{ptsym}), and $M, \alpha, l$ being all real. They proved that for $M>1,$ $\alpha<M+1+|2l+1|$, eigenvalues are all real, and for $M>1,$ $\alpha<M+1-|2l+1|$, they are all positive. A special case of (\ref{bessis}) is the potential $iz^3$ (when $M=\frac{3}{2},\,\alpha=l=0$), which is the $\beta=0$ version  of the BZJ conjecture, but their results do not cover the $\beta\not=0$ version.  (Suzuki \cite{Suzuki} also studied the whole $l=0$ version of (\ref{bessis}) under different boundary conditions.)

The proof of our main theorem, Theorem \ref{main}, has two parts. The first part follows closely the method of Dorey et al.\ \cite{Dorey,Dorey2}, developing functional equations for spectral determinants, expressing them in factorized forms and then studying an ``associated'' eigenvalue problem. We also introduce a symmetry lemma that is required by our more complicated potentials. The second part builds on earlier work of the author in \cite{Shin}, estimating eigenvalues of the ``associated'' problem by integrating over suitably chosen half-lines in the complex plane. Of course both this paper and \cite{Dorey,Dorey2} are indebted to the work of Sibuya \cite{Sibuya}.

Note that our result Corollary \ref{main3} proves the full BZJ conjecture; that is, eigenvalues $\lambda$ of (\ref{zinn}) are all positive real. Also Theorem \ref{main} contains the polynomial potential case ($l=0,\, M\in \N$) of problem (\ref{bessis}), though only with $\alpha\leq 0$, whereas Dorey et al.\ handle $\alpha<M$. (Our proof in the case $\alpha\leq 0$ can be seen to reduce to that of Dorey et al.) In Theorem \ref{exactly_solv} we do manage to handle the case $0<\alpha<M$, by using also the harmonic oscillator inequality, which is a different approach from that used in \cite[page 5701]{Dorey}. 

In a related direction,  Bender and Boettcher \cite{Bender1} found a family of the following quasi-exactly solvable quartic potential problems
\begin{equation}\label{boettcher}
\left[-\frac{d^2}{dz^2}-\left[(iz)^4+2\alpha (iz)^3+(\alpha^2-2\beta)(iz)^2-2(\alpha\beta-J)(iz)\right]\right]u(z)=\lambda u(z)
\end{equation}
with the same boundary conditions as those of (\ref{ptsym}), where $\alpha,\,\beta \in \R$ and $J\in \N$. Note here that the positive integer $J$ denotes the number of the eigenfunctions that can be found exactly in closed form. However, for the purpose of studying the reality of the eigenvalues, we can allow $J\in \R$.
Our results in Theorem \ref{main} confirm that if for any $J\in \R$, we have either $\alpha\, \beta \geq J$ and $\alpha\geq 0$, or $\alpha\, \beta \geq J$ and $2\beta\geq \alpha^2$, then eigenvalues of (\ref{boettcher}) are all positive real.

The above Hamiltonians are not  Hermitian in general. However,  according to Bender and Weniger \cite{Bender5}, Hermiticity of traditional Hamiltonians is a useful mathematical constraint rather than a physical requirement, in order to guarantee real eigenvalues. 
All Hamiltonians mentioned above  are the so-called $\mathcal {PT}$-symmetric Hamiltonians.
A $\mathcal {PT}$-symmetric Hamiltonian is a  Hamiltonian which is invariant under the product of the parity operation $\mathcal P(: z \mapsto -\overline{z})$ (an upper bar denotes the complex conjugate) 
and the time reversal operation $\mathcal T(: i \mapsto
-i)$.   These $\mathcal {PT}$-symmetric Hamiltonians have arisen in recent years in a number of physics papers, see \cite{Savage,Handy2,Handy1,mez,Ali1,Znojil} and other references mentioned above, which support that some $\mathcal {PT}$-symmetric Hamiltonians have real eigenvalues only.
In general the $\mathcal {PT}$-symmetric Hamiltonians are not Hermitian and hence the reality of eigenvalues is not obviously guaranteed. But the important work of Dorey et al.\ \cite{Dorey}, and results in this paper, prove rigorously that some $\mathcal {PT}$-symmetric Hamiltonians indeed have real eigenvalues only.

As a final remark of the introduction, we mention that if $H=-\frac{d^2}{d z^2}+V(z)$ is
$\mathcal{PT}$-symmetric, then $\overline{V(-\overline{z})}=V(z)$ and so $\Re V(z)$ is
an even function and $\Im V(z)$ is an odd function. Hence if $V(z)$ is
a polynomial, then $V(z) =Q(iz)$ for some real polynomial
$Q$. Certainly (\ref{ptsym}) is a $\mathcal {PT}$-symmetric Hamiltonian. 

\section{Properties of the solutions}
\label{properties}
In this section we will introduce some definitions and known facts related with the equation (\ref{ptsym}). One of our main tasks is to identify the eigenvalues as being the zeros of a certain entire function, in Lemma \ref{equiv}. But first, we rotate the equation (\ref{ptsym}) as follows because some known facts, which are related to our argument throughout, are directly available for this rotated equation.

Let $u$ be a solution of (\ref{ptsym}) and let $v(z)=u(-iz)$. Then $v$ solves
\begin{equation}\label{rotated}
-v^\dd(z)+[z^m+P(z)+\lambda]v(z)=0,
\end{equation}
where $m\geq 2$ and $P$  is a real polynomial (possibly, $P\equiv 0$) of the form
$$P(z)=a_1z^{m-1}+a_2z^{m-2}+\cdots +a_{m-1}z.$$

Next we will rotate the boundary conditions. We state them in a more general context by using the following.

\begin{figure}[t]
    \begin{center}
    \includegraphics[width=.4\textwidth]{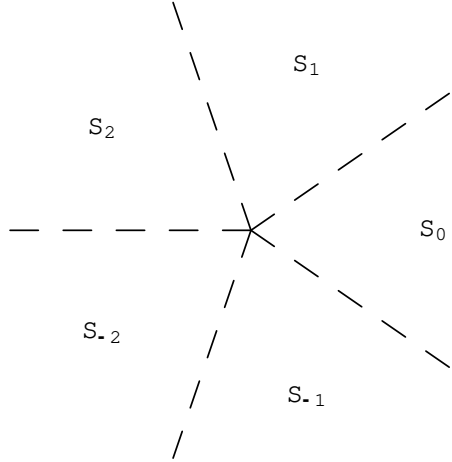}
    \end{center}
 \vspace{-.5cm}
\caption{The Stokes sectors for $m=3$. The dotted rays are $\arg z=\pm\frac{\pi}{5},\,\pm\frac{3\pi}{5},\, \pi.$}\label{f:graph1}
\end{figure}

\begin{definition}
{\rm {\it The Stokes sectors} $S_k$ of the equation (\ref{rotated})
are
$$ S_k=\left\{z\in \C:|\arg z-\frac{2k\pi}{m+2}|<\frac{\pi}{m+2}\right\}\quad\text{for}\quad k\in \Z.$$ }
\end{definition}
\noindent See Figure \ref{f:graph1}.

 It is known that every non-constant solution of (\ref{rotated}) either decays to zero or blows up exponentially, in each Stokes sector $S_k$. Thus the boundary conditions on $u$ in  (\ref{ptsym}) become that $v$ decays in $S_{-1}\cup S_1$.

Before we introduce Sibuya's results, we define  a sequence of complex numbers $b_j$ in terms of the $a_k$ and $\lambda$, as follows. For $\lambda \in \C$ fixed, we expand
\begin{eqnarray}
&&(1+a_1z^{-1}+a_2z^{-2}+\cdots+a_{m-1}z^{1-m}+\lambda z^{-m})^{1/2}\nonumber\\
&=&1+\sum_{k=1}^{\infty}{\frac{1}{2}\choose{k}}\left( a_1z^{-1}+a_2z^{-2}+\cdots+a_{m-1}z^{1-m}+\lambda z^{-m}\right)^k\nonumber\\
&=&1+\sum_{j=1}^{\infty}\frac{b_j}{z^j},\qquad\text{for large}\quad|z|.\label{b_def}
\end{eqnarray}
Note that $b_1,\,b_2,\,\ldots,\,b_{m-1}$ do not depend on $\lambda$. We further define $r_m=-\frac{m}{4}$ if $m$ is odd, and $r_m=-\frac{m}{4}-b_{\frac{m}{2}+1}$ if $m$ is even.

Now we are ready to introduce some existence results  and asymptotic estimates of Sibuya \cite{Sibuya}. The existence of an entire solution with a specified asymptotic representation for fixed $a_k$'s and $\lambda$, is presented as well as an asymptotic expression of the value of the solution at $z=0$ as $\lambda$ tends to infinity. These results are in Theorems 6.1, 7.2 and 19.1 of Sibuya's book \cite{Sibuya}. The following is a special case of these theorems that is enough for our argument later. The coefficient vector $$a:=(a_1,a_2,\ldots, a_{m-1})$$ is allowed to be complex, here. 
\begin{proposition}\label{prop}
The equation (\ref{rotated}), with $a_k\in \C$, $k=1,2,\ldots,m-1$, admits a solution  $f(z,a,\lambda)$ with the following properties.
\begin{enumerate}
\item[(i)] $f(z,a,\lambda)$ is an entire function of $(z,a,\lambda)$.
\item[(ii)] $f(z,a,\lambda)$ and $f^\d(z,a,\lambda)=\frac{d}{dz}f(z,a,\lambda)$ admit the following asymptotic expressions. Let $\epsilon>0$. Then
\begin{eqnarray}
f(z,a,\lambda)&=&\qquad z^{r_m}(1+O(z^{-1/2}))\exp\left[-F(z,a,\lambda)\right],\nonumber\\
f^\d(z,a,\lambda)&=&-z^{r_m+\frac{m}{2}}(1+O(z^{-1/2}))\exp\left[-F(z,a,\lambda) \right],\nonumber
\end{eqnarray}
as $z$ tends to infinity in  the sector $|\arg z|\leq \frac{3\pi}{m+2}-\epsilon$, uniformly on each compact set of $(a,\lambda)$-values . 
Here
\begin{equation}\nonumber
F(z,a,\lambda)=\frac{2}{m+2}z^{\frac{m+2}{2}}+\sum_{1\leq j<\frac{m}{2}+1}\frac{2}{m+2-2j}b_j z^{\frac{1}{2}(m+2-2j)}.
\end{equation}
\item[(iii)] Properties \textup{(i)} and \textup{(ii)} uniquely determine the solution $f(z,a,\lambda)$ of (\ref{rotated}).
\item[(iv)] For each fixed $a$ and $\delta>0$, $f$ and $f^\d$ also admit the asymptotic expressions, 
\begin{eqnarray}
f(0,a,\lambda)&=&[1+o(1)]\lambda^{-1/4}\exp\left[K\lambda^{\frac{1}{2}+\frac{1}{m}}(1+o(1))\right],\label{eq1}\\
f^\d(0,a,\lambda)&=&-[1+o(1)]\lambda^{1/4}\exp\left[K\lambda^{\frac{1}{2}+\frac{1}{m}}(1+o(1))\right],\label{eq2}
\end{eqnarray}
 as $\lambda$ tends to infinity in the sector $|\arg \lambda|\leq \pi-\delta$, where 
\begin{equation}\label{def_K}
 K=\int_0^{\infty}\left(\sqrt{1+t^m}-\sqrt{t^m}\right)\, dt.
\end{equation}
\end{enumerate}
\end{proposition}
\begin{proof}
In  Sibuya's book \cite{Sibuya}, see Theorem 6.1 for a proof of (i) and (ii); Theorem 7.2 for a proof of (iii); and Theorem 19.1 for a proof of (iv). And note that properties (i), (ii) and (iv) are summarized on pages 112--113 of Sibuya's book. 
\end{proof}

We now give references for the proof of Proposition \ref{main2}. We use the number
 $$\omega=\exp\left[\frac{2\pi}{m+2}i\right].$$
\begin{proof}[Proof of Proposition ~\ref{main2}]
See Theorem 29.1 of Sibuya \cite{Sibuya} for a proof which says that eigenvalues are simple, and
\begin{equation}\label{ak}
\lambda_k= \omega^{m}\left(\frac{(-2k+1)\pi}{2K \sin \frac{2\pi}{m}} \right)^{\frac{2m}{m+2}}[1+o(1)],\quad\text{as}\quad k\rightarrow \infty,
\end{equation}
where $K$ is given by (\ref{def_K}). Note that Sibuya studies the equation (\ref{rotated}) with the boundary conditions that $v$ decays in $S_0\cup S_2$, while in this paper we consider the boundary conditions of the rotated equation (\ref{rotated}) that  $v$ decays in $S_{-1}\cup S_1$. The factor $\omega^{m}$ in our formula (\ref{ak}) is due to this rotation of the problem. 

The remaining two claims (I) and (III) are easy consequences of the asymptotic expression (\ref{ak}).

Also one can compute $K$ directly or see the equation (2.22) in \cite{Dorey2}, which says
$$K=-\frac{1}{2\sqrt{\pi}}\Gamma\left(-\frac{1}{2}-\frac{1}{m}\right)\Gamma\left(1+\frac{1}{m}\right).$$
So this along with (\ref{ak}) and the identity $\Gamma(\lambda)\Gamma(1-\lambda)=\pi\csc (\pi \lambda)$ implies (\ref{Bender_exp}).
Note that the asymptotic expression (\ref{Bender_exp}) of the eigenvalues  agrees with that of Bender and Boettcher \cite{Bender} obtained by the WKB calculation for the eigenvalue problem (\ref{Bender_problem}), after an index shift.

We mention that the simplicity of the eigenvalues can be proved by using the fact that for each Stokes sector, there exist two solutions of (\ref{rotated}) with no boundary conditions imposed such that one decays to zero and another blows up as $z$ tends to infinity in the sector.
\end{proof}

The next thing we want to introduce is the Stokes multiplier. First, we let
\begin{equation}\nonumber
G^k(a):=(\omega^{-k}a_1, \omega^{-2k}a_2,\ldots,\omega^{-(m-1)k}a_{m-1})\quad \text{for}\quad k\in \Z.
\end{equation}

Let $f(z,a,\lambda)$ be the function in Proposition \ref{prop}. Note that $f(z,a,\lambda)$ decays to zero exponentially as $z\rightarrow \infty$ in $S_0$ and so $f(z,a,\lambda)$ blows up in $S_{-1}\cup S_1$. Then one can see that the function
$$f_k(z,a,\lambda):=f(\omega^{-k}z,G^k(a),\omega^{-mk}\lambda),$$
 which is obtained by rotating $f(z,G^k(a),\omega^{-mk}\lambda)$, solves (\ref{rotated}). It is also clear that $f_k(z,a,\lambda)$ decays in $S_k$ and blows up in $S_{k-1}\cup S_{k+1}$ since $f(z,G^k(a),\omega^{-mk}\lambda)$ decays in $S_0$. Then since no non-constant solution decays in two consecutive Stokes sectors, $f_{k}$ and $f_{k+1}$ are linearly independent and hence any solution of (\ref{rotated}) can be expressed as a linear combination of these two. Especially, for some coefficients $C(a,\lambda)$ and $\widetilde{C}(a,\lambda)$,
\begin{equation}\label{stokes}
f_{-1}(z,a,\lambda)=C(a,\lambda)f_0(z,a,\lambda)+\widetilde{C}(a,\lambda)f_{1}(z,a,\lambda).
\end{equation}
These $C(a,\lambda)$ and $\widetilde{C}(a,\lambda)$ are called {\it the Stokes multipliers of $f_{-1}$ with respect to $f_0$ and $f_1$}.

We then see that 
$$C(a,\lambda)=\frac{W_{-1,1}(a,\lambda)}{W_{0,1}(a,\lambda)}\quad\text{and}\quad \widetilde{C}(a,\lambda)=-\frac{W_{-1,0}(a,\lambda)}{W_{0,1}(a,\lambda)},$$
where $W_{j,k}=f_jf_k^\d -f_j^\d f_k$ is the Wronskian of $f_j$ and $f_k$. Since both $f_j,\,f_k$ are solutions of the same linear equation (\ref{rotated}), we know that the Wronskians are constant functions of $z$. Since $f_k$ and $f_{k+1}$ are linearly independent, $W_{k,k+1}\not=0$ for all $k\in \Z$. Moreover, we have the following which is needed in the proof of our main theorem. 

\begin{lemma}\label{unit}
The Stokes multiplier $\widetilde{C}(a,\lambda)$ is independent of $\lambda$. Moreover,
if $a \in \R^{m-1}$ then $|\widetilde{C}(a,\lambda)|=1$.
\end{lemma}
\begin{proof}
First note that Sibuya's multiplier $\widetilde{c}(a,\lambda)$ in \cite{Sibuya} is $W_{1,0}/W_{1,2}$ while we use $\widetilde{C}(a,\lambda)=W_{-1,0}/W_{0,1}$. Since $f_k(z,\,a,\,\lambda)=f(\omega^{-k}z,G^k(a),\omega^{-mk}\lambda)$, we see that
\begin{eqnarray}
 f_{k+1}(z,\,a,\,\lambda)
&=&f(\omega^{-(k+1)}z,G^{k+1}(a),\omega^{-m(k+1)}\lambda)\nonumber\\
&=&f_k(\omega^{-1}z,G(a),\omega^{-m}\lambda).\nonumber
\end{eqnarray}
Hence using $\omega^{m+2}=1$, we see that 
\begin{equation}\label{kplus1}
W_{k+1,j+1}(a,\lambda)=\omega^{-1}W_{k,j}(G(a),\omega^2\lambda),
\end{equation}
 which is the equation (26.28) of \cite{Sibuya}.

So using the equation (26.29) on page 117 of \cite{Sibuya}, one can get
\begin{eqnarray}
\widetilde{C}(a,\lambda)&=& -\frac{W_{-1,0}(a,\lambda)}{W_{0,1}(a,\lambda)}\nonumber\\
                        &=& -\frac{W_{0,1}(G^{-1}(a),\omega^{-2}\lambda)}{W_{1,2}(G^{-1}(a),\omega^{-2}\lambda)},\qquad\text{by (\ref{kplus1}),}\nonumber\\
                        &=&-\omega^{1-2\widetilde{\nu}(G^{-1}(a))},\qquad\text{by (26.29) of \cite{Sibuya}},\label{eq121}
\end{eqnarray}
where
\begin{eqnarray}
\widetilde{\nu}(G^{-1}(a))=\left\{
                         \begin{array}{rl}
                         0 & \quad \text{if $m$ is odd,}\\
                         b_{m/2+1}(G^{-1}(a))& \quad \text{if $m$ is even.}
                          \end{array}
                         \right. \nonumber
\end{eqnarray}
From (\ref{eq121}), it is clear that $\widetilde{C}(a,\lambda)$ is independent of $\lambda$.
We want $\widetilde{\nu}(G^{-1}(a))$ to be real if $a \in \R^{m-1}$, so that $|\widetilde{C}(a,\lambda)|=|\omega|=1$. Suppose $a \in \R^{m-1}$.
Since $b_{\frac{m}{2}+1}(G^{-1}(a))=-b_{\frac{m}{2}+1}(a)$  as noted on  page 117 of \cite{Sibuya} (or can be directly verified from (\ref{b_def})), it is sufficient to show that $b_{\frac{m}{2}+1}(a)$ is real when $a\in \R^{m-1}$. Since $a_k$'s are all real,  from (\ref{b_def}) we conclude that $b_{\frac{m}{2}+1}(a)$ must be real. This completes the proof.
\end{proof}

Thus from the proof of Lemma \ref{unit} we get $\widetilde{C}(a,\lambda)=e^{i\phi_0}$ for some $\phi_0= \phi_0(a)\in \R$ and hence from (\ref{stokes}) we have
\begin{eqnarray}
C(a,\lambda)f_0(z,a,\lambda)&=&f_{-1}(z,a,\lambda)-e^{i\phi_0}f_1(z,a,\lambda)\label{dependent}\\
                            &=&f(\omega z, G^{-1}(a),\omega^m \lambda)-e^{i\phi_0}f(\omega^{-1}z,G(a),\omega^{-m}\lambda).\label{relation}
\end{eqnarray}
From this, for each $a\in \R^{m-1}$ we can relate the zeros of $C(a,\lambda)$ with the eigenvalues of (\ref{ptsym}) as follows.
\begin{lemma}\label{equiv}
For each  fixed $a=(a_1,a_2,\ldots,a_{m-1})\in \R^{m-1}$,
a complex number $\lambda$ is an eigenvalue of (\ref{ptsym}) if and only if  $\lambda$ is a zero of the entire function $C(a,\lambda)$. 
\end{lemma}
Hence, the eigenvalues are discrete because they are zeros of a non-constant entire function.
Note that the Stokes multiplier $C(a,\lambda)$ is called a {\it spectral determinant} or an {\it Evans function}, because its zeros are all eigenvalues of an eigenvalue problem. 
\begin{proof}
Suppose that $\lambda$ is an eigenvalue of (\ref{ptsym}) with the corresponding eigenfunction $u$. Then we let $v(z)=u(-iz)$, and hence $v$ solves (\ref{rotated}) and decays in $S_{-1}\cup S_1$. 
Since $f_{-1}$ is another solution of (\ref{rotated}) that decays in $S_{-1}$, we see that $f_{-1}$ is a multiple of $v$. Similarly $f_1$ is a multiple of $v$. Hence the right-hand side of (\ref{dependent}) decays in $S_{-1}\cup S_1$. But  $f_0$ blows up in $S_{-1}\cup S_1$, and so (\ref{dependent}) implies $C(a,\lambda)=0$.

Conversely we suppose that  $C(a,\lambda)=0$ for some $\lambda\in \C$. Then
from (\ref{dependent}) we see that $f_{-1}$ is a constant multiple of $f_{1}$. Thus both are decaying in $S_{-1}\cup S_1$ and hence $u(z):=f_{-1}(iz,a,\lambda)$ is an eigenfunction of (\ref{ptsym}) with the corresponding eigenvalue $\lambda$.
\end{proof}

Next we examine (\ref{relation}) and its differentiated form at $z=0$, which are,
\begin{eqnarray}
C(a,\lambda)f(0,a,\lambda)&=&f(0,G^{-1}(a),\omega^{m}\lambda)-e^{i\phi_0}f(0,G(a),\omega^{-m}\lambda),\label{eq3}\\
C(a,\lambda) f^\d(0,a,\lambda)&=&\omega f^\d(0,G^{-1}(a),\omega^{m}\lambda) -e^{i\phi_0}\omega^{-1}f^\d(0,G(a),\omega^{-m}\lambda).\label{eq4}
\end{eqnarray}
The right-hand sides of these are given by differences of two functions of $\lambda$. We will express these right-hand sides with single functions, respectively. To this end, we prove that $f$ and $f^\d$ both have some symmetry as follows.
\begin{lemma}
Let ${a}=({a}_1,{a}_2,\ldots,{a}_{m-1})\in \C^{m-1}$.  Then we have 
\begin{equation}\label{eq5}
f(0,{a},\lambda)=\overline{f(0,\overline{{a}},\overline{\lambda})}\quad\text{and}\quad
f^\d(0,{a},\lambda)=\overline{f^\d(0,\overline{{a}},\overline{\lambda})}.
\end{equation}
Especially, we have that if $a=({a}_1,{a}_2,\ldots,{a}_{m-1})\in \R^{m-1}$ is real, then
\begin{equation}\label{eq7}
f(0,G(a),\lambda)=\overline{f(0,G^{-1}(a),\overline{\lambda})}\quad\text{and}\quad
f^\d(0,G(a),\lambda)=\overline{f^\d(0,G^{-1}(a),\overline{\lambda})}.
\end{equation}
\end{lemma}
\begin{proof}
Let $g(z)=f(z,{a},\lambda)$, which is the entire function $f$ in Proposition \ref{prop} and hence decays in $S_0$. Then $g$ solves
$$-g^\dd(z)+(z^m+{a}_1z^{m-1}+{a}_2z^{m-2}+\cdots+{a}_{m-1}z+\lambda)g(z)=0.$$
Next we take the complex conjugate of this and replace $z$ by $\overline{z}$. Then we see that $\overline{g(\overline{z})}$ is entire and solves the following equation
\begin{equation}\label{eq10}
-\overline{g^\dd(\overline{z})}+(z^m+\overline{{a}_1}z^{m-1}+\overline{{a}_2}z^{m-2}+\cdots+\overline{{a}_{m-1}}z+\overline{\lambda})\overline{g(\overline{z})}=0.
\end{equation}
Since the entire functions $\overline{g(\overline{z})}$ and $f(z,\overline{{a}},\overline{\lambda})$ are  solutions of (\ref{eq10}) that decay in $S_0$, we see that these two are linearly dependent. So one is a constant multiple of the other. Moreover, from  (\ref{b_def})  we see that 
$\overline{b_k(a,\lambda)}=b_k(\overline{a},\overline{\lambda})$ for all $k\in \N$
where we used $b_k(a,\lambda)$ instead of $b_k$ to indicate its dependence on $a$ and $\lambda$.
Also we have $\overline{F(\overline{z},a,\lambda)}=F(z,\overline{a},\overline{\lambda})$ in Proposition \ref{prop}. Hence the entire functions $\overline{g(\overline{z})}$ and $f(z,\overline{a},\overline{\lambda})$ along with their first derivatives satisfy the same asymptotic expressions in  Proposition \ref{prop} (ii), so we conclude that 
\begin{equation}\label{eq111}
\overline{g(\overline{z})}=f(z,\overline{a},\overline{\lambda})
\end{equation}
 by  Proposition \ref{prop} (iii). Next substituting $z=0$ in (\ref{eq111}) gives the first equation in (\ref{eq5}). Also we differentiate (\ref{eq111}) with respect to $z$ and substitute $z=0$ to get the second equation in (\ref{eq5}).
For (\ref{eq7}), just note that $\overline{G(a)}=G^{-1}(\overline{a}).$
\end{proof}
Next we want infinite product representations of $f(0,a,\lambda)$ and 
$f^\d(0,a,\lambda)$, with respect to $\lambda$. But first, we recall the definition of  order of an entire function, which will be needed in the proof of the next lemma.
Let $M(r, g)=\max \{|g(re^{i\theta})|: 0\leq \theta\leq 2\pi\}$ for $r>0$. Then the order of an entire function $g$ is 
$$\limsup_{r\rightarrow \infty}\frac{\log \log M(r,g)}{\log r}.$$
If for some positive real numbers $\sigma,\, c_1,\, c_2$, we have $M(r,g)\leq c_1 \exp[c_2 r^{\sigma}]$ for all large $r$, then the order of $g$ is finite and less than or equal to $\sigma$. 
\begin{lemma}\label{asymp}
Suppose $m\geq 3$. The functions $\lambda \mapsto f(0,a,\lambda)$ and  $\lambda \mapsto f^\d(0,a,\lambda)$ have infinitely many zeros $E_j$ and $E^\d_j$, respectively. They admit the following infinite product representations for each fixed $a=(a_1,a_2,\ldots,a_{m-1})\in \C^{m-1}$:
\begin{eqnarray}
f(0,a,\lambda)&=&D_0\lambda^{n_0}\prod_{j=1}^{\infty}\left(1-\frac{\lambda}{E_j}\right)\quad\text{for some  $D_0\in \C$ and nonnegative integer $n_0$,}\nonumber\\
f^\d(0,a,\lambda)&=&D_1\lambda^{n_1}\prod_{j=1}^{\infty}\left(1-\frac{\lambda}{E^\d_j}\right)\quad\text{for some  $D_1\in \C$ and nonnegative integer $n_1.$}\nonumber
\end{eqnarray}
Moreover, these infinite products converge absolutely.
\end{lemma}
\begin{proof}
If we show that both $f(0,a,\lambda)$ and  $f^\d(0,a,\lambda)$ have orders (with respect to $\lambda$) strictly less than one, then this lemma is a consequence of the Hadamard factorization theorem (see, for example, Theorem 14.2.6 on page 199 of \cite{Hille2}). So we will show that  $f(0,a,\lambda)$ and  $f^\d(0,a,\lambda)$ have orders strictly less than one.

From the equations  (\ref{eq1}) and (\ref{eq2}), we see that except for $\pi-\delta \leq \arg \lambda \leq \pi+\delta$, small $\delta>0$, both $|f(0,a,\lambda)|$ and  $|f^\d(0,a,\lambda)|$ are bounded by $\exp[2K|\lambda|^{\frac{m+2}{2m}}]$ for large $|\lambda|$. So to show that they have the orders strictly less than one, it  suffices to show that for $\pi-\delta \leq \arg \lambda \leq \pi+\delta$, small $\delta>0$, they are bounded by $d_1\exp[d_2|\lambda|^{\frac{m+2}{2m}}]$ for some $d_1>0$ and $d_2>0$.

From (\ref{stokes}) with $z=0$, one can see that
$$\left|f(0,G^{-1}(a),\lambda)\right|\leq\left|C(a,\omega^2\lambda)f(0,a,\omega^2\lambda)\right|+\left|\widetilde{C}(a,\omega^2\lambda)f(0,G(a),\omega^{-4}\lambda)\right|.$$
In this inequality, let $\lambda$ lie in the region $|\arg \lambda-\pi| \leq\delta$. Then since $\omega^2\lambda$ and $\omega^{-4}\lambda$ are not in  $|\arg \lambda-\pi| \leq\delta$, we can use the asymptotic expression (\ref{eq1}) to get that for all large $|\lambda|$,
$$\left|f(0,G^{-1}(a),\lambda)\right|\leq \left[\left|C(a,\omega^2\lambda)\right|+\left|\widetilde{C}(a,\omega^2\lambda)\right|\right]\exp[2K |\lambda|^{\frac{m+2}{2m}}].$$

We know that $\left|\widetilde{C}(a,\omega^2\lambda)\right|$ depends only on $a$ by Lemma \ref{unit}. Also the equations (29.4) and (29.7)  imply that for fixed $a$, 
$$\left|C(a,\omega^2\lambda)\right|\leq d_3\exp[2K|\lambda|^{\frac{m+2}{2m}}],\qquad\text{for some $d_3>0$}.$$
Thus we see that for each $a$,
$$\left|f(0,G^{-1}(a),\lambda)\right|\leq d_1\exp[4|\lambda|^{\frac{m+2}{2m}}],\qquad \text{for some $d_1>0$.}$$
Hence the order of $f(0,a,\lambda)$ with respect to the $\lambda$-variable is less than or equal to $\frac{m+2}{2m}$. Hence by combining this with (\ref{eq1}), we conclude that the order of $f(0,a,\lambda)$ is $\frac{m+2}{2m}$, which is strictly less than one. 

Next we differentiate (\ref{stokes}) with respect to the $z$-variable and set $z=0$, then similarly using (\ref{eq2}),  we can conclude that the order of  $f^\d(0,a,\lambda)$ is $\frac{m+2}{2m}$.
\end{proof}

\section{Proof of Theorem \ref{main}}\label{main_proof}
When $m=2$, the equation (\ref{ptsym}) is a translation of the harmonic oscillator. So there is nothing new here. We mention that since $z^2+a_1i z=(z+\frac{a_1}{2}i)^2+\frac{a_1^2}{4}$, the eigenvalues for the potential  $z^2+a_1iz$ are $2k+1+\frac{a_1^2}{4}>0$.

Suppose $m\geq 3$ and suppose that $\lambda\in \C$ is an eigenvalue of the eigenproblem (\ref{ptsym}), then by Lemma \ref{equiv} we have $C(a,\lambda)=0$. Then from (\ref{eq3}) and (\ref{eq4}) along with (\ref{eq7}), we have 
\begin{eqnarray}
0&=&f(0,G^{-1}(a),\omega^{m}\lambda)-e^{i\phi_0}\overline{f(0,G^{-1}(a),\omega^{m}\overline{\lambda})},\nonumber\\
0&=&\omega f^\d(0,G^{-1}(a),\omega^{m}\lambda) -e^{i\phi_0}\omega^{-1}\overline{f^\d(0,G^{-1}(a),\omega^{m}\overline{\lambda})}.\nonumber
\end{eqnarray}
Since the non-constant function $f(z,G^{-1}(a),\omega^{m}\lambda)$ solves a linear second order ordinary differential equation, both $f(0,G^{-1}(a),\omega^{m}\lambda)$ and $f^\d(0,G^{-1}(a),\omega^{m}\lambda)$ cannot be zero at the same time; otherwise, $f(z,G^{-1}(a),\omega^{m}\lambda)\equiv 0$.

Suppose that $f(0,G^{-1}(a),\omega^{m}\lambda)\not=0$.
Then from Lemma \ref{asymp} we have
$$0\not=D_0(\omega^m\lambda)^{n_0}\prod_{j=1}^{\infty}\left(1-\frac{\omega^m\lambda}{E_j}\right)= e^{i\phi_0}\overline{D_0(\omega^m\overline{\lambda})^{n_0}\prod_{j=1}^{\infty}\left(1-\frac{\omega^m\overline{\lambda}}{E_j}\right) }.$$
Then  by equating the absolute values of the two sides of the equation (and using $\omega^{m+2}=1$),  we have 
\begin{equation}\label{product_form}
\prod_{j=1}^{\infty}\left|\frac{\omega^2E_j-\lambda}{\omega^2E_j-\overline{\lambda}}\right|=1.
\end{equation}
Likewise, when  $f^\d(0,G^{-1}(a),\omega^{m}\lambda)\not=0$, we get the following. 
\begin{equation}\label{product_form2}
\prod_{j=1}^{\infty}\left|\frac{\omega^2E^\d_j-\lambda}{\omega^2E^\d_j-\overline{\lambda}}\right|=1.
\end{equation}

We mention that $\omega^2E_j$ and  $\omega^2E^\d_{j^\d}$ lie in the open lower half-plane for some $j,\,j^\d$.
 From Lemma \ref{asymp} we know that $f(0,G^{-1}(a),E)$ and $f^\d(0,G^{-1}(a),E)$ have infinitely many zeros $E_*$. And  (\ref{eq1}) and (\ref{eq2})  imply  that the zeros $E_*$ near infinity lie near the negative real axis. Thus certainly $\Im \omega^2E_j<0$ and  $\Im \omega^2E^\d_{j^\d}<0$ for some $j,\,j^\d$.

Below we will show that the hypotheses on the signs of the coefficients $a_1,a_2,\ldots,a_{m-1}$ of $P$ force all the $\omega^2E_j$ and $\omega^2E^\d_j$ to lie in the closed lower half-plane, which implies either
\begin{eqnarray}
&& \left|{\omega^2E_j-\lambda}\right|\geq \left|{\omega^2E_j-\overline{\lambda}}\right|\quad\text{and}\quad \left|{\omega^2E^\d_j-\lambda}\right|\geq\left|{\omega^2E^\d_j-\overline{\lambda}}\right|,\quad\text{$\forall j\in \N,\,$ if $\,\Im \lambda\geq 0,$ or}\label{eq*}\\
&&\left|{\omega^2E_j-\lambda}\right|\leq  \left|{\omega^2E_j-\overline{\lambda}}\right|\quad\text{and}\quad\left|{\omega^2E^\d_j-\lambda}\right|\leq \left|{\omega^2E^\d_j-\overline{\lambda}}\right|,\quad\text{$\forall j\in \N,\,$ if $\,\Im \lambda\leq 0,$} \nonumber
\end{eqnarray}
since $\lambda$ and $\overline{\lambda}$ are reflections of each other with respect to the real axis.
If (\ref{product_form}) holds, then (\ref{eq*}) implies
$\left|{\omega^2E_j-\lambda}\right|=\left|{\omega^2E_j-\overline{\lambda}}\right|$ for all $j\in \N$. If (\ref{product_form2}) holds, then (\ref{eq*}) implies $\left|{\omega^2E^\d_j-\lambda}\right|=\left|{\omega^2E^\d_j-\overline{\lambda}}\right|$ for all $j\in \N$.
Since $\Im \omega^2 E_j < 0$ for some $j$ and $\Im \omega^2 E^\d_j <0$ for some $j$, and since $\lambda$ and $\overline{\lambda}$ are reflections of each other with respect to the real axis, we deduce in either case that
$\lambda=\overline{\lambda}$ and hence $\lambda$ is real. 

So our next task is to show that all the $\omega^2E_j$ and $\omega^2E^\d_j$ lie in  the closed lower half-plane.
Suppose that for some $E_{*} \in \C$,
\begin{equation}\label{eq_33}
\text{either}\quad f(0,G^{-1}(a),E_{*})=0\quad\text{or} \quad f^\d(0,G^{-1}(a),E_{*})=0.
\end{equation}
That is, either $E_*=E_j$ or  $E_*=E^\d_j$ for some $j\in \N$. We know that $v(z)=f(z,G^{-1}(a),E_{*})$ solves
\begin{equation}\label{eq20}
-v^\dd(z)+\left[z^m+\sum_{k=1}^{m-1} a_k\omega^k z^{m-k} \right]v(z)=-E_{*}v(z),
\end{equation}
where $a_k\in \R,\,\,k=1,2,\ldots,m-1,$ and by  (\ref{eq_33}), $v$ satisfies either Dirichlet ($E_*=E_j$) or Neumann boundary ($E_*=E_j^\d$) condition at $0$, and Dirichlet condition at $\infty+0i$. We call   (\ref{eq20}) with these boundary conditions the ``associated'' eigenvalue problem. We aim to show all the eigenvalues $E_*$ have $\Im (\omega^2 E_*)\leq 0$.  

Let $g(r)=v(re^{i\theta})$ with $\theta$ fixed, $|\theta|<\frac{\pi}{m+2}$. We then multiply (\ref{eq20}) by  $\omega^2e^{-2i\theta}\overline{g(r)}$ and integrate over $0\leq r<\infty$ to get 
\begin{eqnarray}
&&-\omega^2e^{-2i\theta}\int_0^{\infty}g^\dd(r)\overline{g(r)}\, dr
+\int_0^{\infty}\left[\omega^2e^{mi\theta}r^m+\sum_{k=1}^{m-1}\left(a_k\omega^{k+2} e^{(m-k)i\theta}r^{m-k}\right)\right]|g(r)|^2\,dr \label{integrable}\\
&=&-\omega^2E_{*} \int_0^{\infty}|g(r)|^2\,dr.\nonumber
\end{eqnarray}
Since $f(z,G^{-1}(a),E_{*})$ decays to zero exponentially in $S_0$, we know the integrability of every term in (\ref{integrable}) for each $|\theta|<\frac{\pi}{m+2}$.
Next we  integrate the first term by parts, using $g(0)=0$ or $g^\d(0)=0$ by (\ref{eq_33}), so the boundary term vanishes. And then taking the imaginary part of the resulting equation gives
\begin{eqnarray}
&&\sin\left(\frac{4\pi}{m+2}-2\theta\right) \int_0^{\infty}|g^\d(r)|^2\, dr+\sin\left(m\theta+ \frac{4\pi}{m+2}\right)\int_0^{\infty}r^m|g(r)|^2\, dr\nonumber\\
&+&\sum_{k=1}^{m-1}a_k\sin\left(\frac{2(k+2)\pi}{m+2}+(m-k)\theta\right)\int_0^{\infty} r^{m-k} |g(r)|^2\,dr\nonumber\\
&&\qquad\qquad=- \Im \left(\omega^2E_{*} \right)\int_0^{\infty}|g(r)|^2\, dr.\label{im_part}
\end{eqnarray} 
 
Recall our hypothesis that $(j-k)a_k\geq 0$ for all $1\leq k\leq m-1$ for some $1\leq j\leq \frac{m}{2}.$ We want to prove the reality of the eigenvalues by showing that $\Im \omega^2E_*\leq 0$ for all the $E_*$. To this end,
we will divide the proof into two cases; {\it Case I}, when $1\leq j \leq \frac{m}{2}$ and $m\geq 5$, or  when $j=1$ and $m=3,\,4$; and  {\it Case II}, when $j=2$ and $m=4$.
  
{\it Case I:} when $1\leq j \leq \frac{m}{2}$ and $m\geq 5$, or when $j=1$ and $m=3,\, 4$.
We choose $\theta$ in (\ref{im_part}) by
\begin{equation}\label{theta_j}
\theta=\frac{(m-2j-2)\pi}{(m-j)(m+2)},
\end{equation}
where the motivation for this choice will be fairly clear later in the proof.
Notice here that $|\theta|<\frac{\pi}{m+2}$ as required.
Then
\begin{eqnarray}
0< \frac{4\pi}{m+2} -2\theta=\frac{2\pi}{m-j}\leq \pi,\quad &\text{ and  hence}&\quad \sin\left(\frac{4\pi}{m+2} -2\theta\right)\geq 0,\quad\text{and}\nonumber\\
0\leq m\theta+ \frac{4\pi}{m+2} =\frac{(m-2j)\pi}{m-j}<\pi,\quad &\text{ and  hence}&\quad \sin\left(m\theta+\frac{4\pi}{m+2} \right)\geq 0.\nonumber
\end{eqnarray}
(Clearly these inequalities use that $j\leq \frac{m}{2}$.)
Also we see that
$$
\frac{2(k+2)\pi}{m+2}+(m-k)\theta =\left(1-\frac{j-k}{m-j}\right)\pi,\quad\text{for all $k$}, 
$$
and so 
$$\sin\left( 1-\frac{j-k}{m-j} \right)\pi \quad\text{has the same sign as $(j-k)$ for all $1\leq k\leq m-1$}.$$
Among other things, this is  why we choose the $\theta$ as above.
So from (\ref{im_part}) and the hypothesis $(j-k)a_k\geq 0$, we conclude that
$\Im \left(\omega^2 E_{*}\right)\leq 0.$
This proves that the eigenvalue $\lambda$ is real.

{\it Case II:} when  $j=2$ and $m=4$.
The reason we separate this case from {\it Case I} is that in this case, $|\theta|=\frac{\pi}{6}=\frac{\pi}{m+2}$, whereas our argument needed $|\theta|<\frac{\pi}{m+2}$ in order to get
 integrability for  terms in (\ref{integrable}). So we modify the proof as follows. For  $\epsilon>0$ small, we multiply (\ref{integrable}) by $e^{-2i\epsilon}$ and set $\theta=-\frac{\pi}{6}+\epsilon$. Then integrating the first term by parts and taking the imaginary part of the resulting equation give
\begin{eqnarray}
&&\sin (4\epsilon) \int_0^{\infty}|g^\d(r)|^2\, dr+\sin (2\epsilon)\int_0^{\infty}r^4|g(r)|^2\, dr\nonumber\\
&+&\sum_{k=1}^{3}a_k\sin\left(\frac{k\pi}{2}+(2-k)\epsilon \right)\int_0^{\infty} r^{4-k} |g(r)|^2\,dr\label{lhs}\\
&&=- \Im \left(\omega^2E_{*} e^{-2i\epsilon}\right)\int_0^{\infty}|g(r)|^2\, dr.\nonumber
\end{eqnarray} 
Clearly $\sin\left(\frac{k\pi}{2}+(2-k)\epsilon \right)$ has the same sign as $(2-k)$ for $1\leq k\leq 3$.
Using the hypothesis $(2-k)a_k\geq 0$ for all $1\leq k\leq 3$, we have that the left side of (\ref{lhs}) is nonnegative and so
$$\Im \omega^2E_* e^{-2i\epsilon}\leq 0.$$
Thus by sending $\epsilon$ to zero, we get
$$\Im \omega^2 E_*\leq 0,$$
which proves the reality for the case of $j=2$ and $m=4$.

Therefore, the eigenvalues of (\ref{ptsym}) are all real under the hypotheses on the $a_k$'s given in the statement of this theorem.

We must still prove the positivity of the eigenvalues. Suppose $u$ is an eigenfunction of (\ref{ptsym}) with an eigenvalue $\lambda\in \R$, and suppose $a_k$'s satisfy the hypotheses of the theorem. Let $v(z)=u(-iz)$. Then we have the equation (\ref{rotated}) with the boundary 
conditions that $v$ decays in 
$$S_{-1}\cup S_1=\{z\in \C:\frac{\pi}{m+2}<|\arg z|<\frac{3\pi}{m+2}\}.$$
Since $\lambda$ and all $a_k$'s are real, one can see that $\overline{v(\overline{z})}$ satisfies the same equation and decays in $S_{-1}\cup S_1$. Then since the eigenvalues are simple, $v(z)$ and $\overline{v(\overline{z})}$ must be linearly dependent, and hence  $v(z)=c\overline{v(\overline{z})}$ for some $c\in \C$. Since  $|v(z)|$ and $|v(\overline{z})|$ agree on the real line, we see that $|c|=1$ and so $|v(z)|=|v(\overline{z})|$ for all  $z\in \C.$ That is, $|v(x+iy)|$ is even in $y$.
From this we have that 
\begin{equation}\label{real=0}
0=\left.\frac{\partial}{\partial\,y}|v(x+iy)|^2\right|_{y=0}=-2\Im \left(v^\d(x)\overline{v(x)}\right),\quad\text{for all}\quad x\in \R.
\end{equation}

Next we let $h(r)=v(r e^{i\theta})$. 
By substituting into the differential equation (\ref{rotated}), then multiplying by $\overline{h(r)}$ and integrating, we get
\begin{eqnarray}
&&-\int_0^{\infty}h^\dd(r)\overline{h(r))}\, dr
+\int_0^{\infty}\left[e^{(m+2)i\theta}r^m+\sum_{k=1}^{m-1} a_k e^{(m-k+2)i\theta}r^{m-k} \right]|h(r)|^2\,dr \nonumber\\
&=&-\lambda e^{2i\theta} \int_0^{\infty}|h(r)|^2\,dr,\quad\text{for}\quad \frac{\pi}{m+2}<\theta<\frac{3\pi}{m+2}.\nonumber
\end{eqnarray}
Integrating the first term by parts and using $h^\d(0)= e^{i\theta}v^\d(0)$, one can get 
\begin{eqnarray}
&-&v^\d(0)\overline{v(0)}+e^{-i\theta}\int_0^{\infty}|h^\d(r)|^2\, dr
+\int_0^{\infty}\left[e^{(m+1)i\theta}r^m+\sum_{k=1}^{m-1} a_k e^{(m-k+1)i\theta}r^{m-k} \right]|h(r)|^2\,dr \nonumber\\
&&=-\lambda e^{i\theta} \int_0^{\infty}|h(r)|^2\,dr,\quad\text{for}\quad \frac{\pi}{m+2}<\theta<\frac{3\pi}{m+2}.\nonumber
\end{eqnarray}
Taking the imaginary part and using (\ref{real=0}) at $x=0$, we have 
\begin{eqnarray}
&&\sin\theta\int_0^{\infty}|h^\d|^2\,dr- \int_0^{\infty}\left[r^m\sin(m+1)\theta +\sum_{k=1}^{m-1} a_k r^{m-k} \sin(m-k+1)\theta \right]|h|^2\,dr\nonumber\\
&&=\lambda \sin \theta \int_0^{\infty}|h|^2\,dr,\quad\text{for all}\quad \frac{\pi}{m+2}<\theta<\frac{3\pi}{m+2}.\label{sine-value}
\end{eqnarray}
(Here again we used that $\lambda$ is real.)
We choose 
$$\theta =\frac{\pi}{m-j+1},$$
so that 
$$\frac{\pi}{m+2}<\frac{\pi}{m+1}\leq\theta\leq\frac{2\pi}{m+1}<\frac{3\pi}{m+2}<\pi$$
as required, and 
$$\sin(m-k+1)\theta=\sin\left(\frac{m-k+1}{m-j+1}\pi \right)$$
has the same sign as $(k-j)$, for all $k$.
 Since  $(k-j)a_k\leq 0$, $\sin\theta\geq 0$ and $\sin(m+1)\theta\leq 0$, we see that the left-hand side of (\ref{sine-value}) is positive, and hence so is the right-hand side. Therefore, the real number $\lambda$ must be positive. This completes the proof of Theorem \ref{main}.

\begin{remarks}$\left.\right.$

{\rm 1. The idea of using the infinite product in (\ref{product_form}) to prove reality of the eigenvalues is due to Dorey et al.\cite{Dorey}. But their potentials are much simpler and the $E_*$ are all negative real in their situation, so that (\ref{eq*}) is immediate. Here is not.}

{\rm 2. The ideas above for proving positivity of the eigenvalues are similar to those used earlier by the author in \cite{Shin}.}

{\rm 3. We note that the hypotheses assumed in Theorem \ref{main} on the coefficients of $P$ are sufficient for real eigenvalues, but not necessary, for at least two reasons. Let $Q(z)=-[z^m+P(z)]$. Then
first, the problem (\ref{ptsym}) with the potential $Q(iz)=-[(iz)^3-(iz)^2]$ is covered by Theorem \ref{main} while the problem  with $Q_1(iz)=-[(iz)^3+2(iz)^2+iz]$ is not. However, $Q(z+1)=Q_1(z)$ and so the potential $Q_1(iz)$ produces positive real eigenvalues only. For general cases, for a real polynomial $Q$ if the problem (\ref{ptsym}) with the potential $Q(iz)$ has positive real eigenvalues $\lambda$ only, then the problem with the potential $Q(iz+c)-Q(c)$ for some real $c\in \R$ has eigenvalues $\lambda-Q(c)$ which are all real.  Second, in the proof of  Theorem \ref{main} in order to ensure that $\Im \omega^2E_*\leq 0$,  we insisted that each and every term on the left-hand side of (\ref{im_part}) has a single sign, and it is clear from Section \ref{extension_sec} below that this is not  necessary.}
\end{remarks}
\section{Extensions of Theorem \ref{main}}
\label{extension_sec}
In this section, we study two particular classes of polynomial potentials to illustrate different methods for sharpening Theorem \ref{main}.
\begin{theorem}\label{ext_thm}
Let $m\geq 4$ and suppose $\alpha <0$,  $\gamma <0$. Suppose that an entire function $u$ along with $\lambda\in \C$ solves
the equation (\ref{ptsym}) with $P(z)=\alpha z^3+\beta z^2+\gamma z$. Then the eigenvalue $\lambda$ is positive real, provided that 
\begin{equation}\label{eq51}
\beta\leq \sqrt{\alpha \gamma}\sqrt{3-\tan^2\left(\frac{\pi}{m}\right)}.
\end{equation} 
The eigenvalue $\lambda$ is also positive real provided that $\lambda\in \R$ and
\begin{equation}\label{eq52}
\beta\leq 4\sqrt{2}\sqrt{\alpha \gamma}\frac{\sqrt{1-\tan^2\left(\frac{\pi}{m+1}\right)}}{3-\tan^2\left(\frac{\pi}{m+1}\right)}.
\end{equation}
\end{theorem}
\begin{remarks}
{\rm ${}$

1. Note that for $\alpha,\,\beta,\,\gamma\leq 0$,  we have $\lambda >0$ by Theorem \ref{main}. The point of Theorem \ref{ext_thm}, then, is that if $\alpha,\,\gamma<0$ then we can allow some values of $\beta>0$.

2. The right-hand side of (\ref{eq51}) is less than that of (\ref{eq52}) as we show at the end of the proof.}
\end{remarks}

\begin{proof}[Proof of Theorem ~\ref{ext_thm}] 
Since the theorem for $\beta\leq 0$ is contained in Theorem \ref{main}, it suffices to show the claims of the theorem hold  under the hypotheses (\ref{eq51}) and (\ref{eq52}) with $\beta$ replaced by $|\beta|$. In proving this we will closely follow the proof of Theorem \ref{main}.

As in the proof of Theorem \ref{main}, in order to prove the reality of the eigenvalues we show that $\Im (\omega^2 E_*)\leq 0$ for all $E_*\in \C$ satisfying (\ref{eq_33}).  In this case, we see that (\ref{im_part})  becomes
\begin{eqnarray}
&&\sin\left(2\phi\right) \int_0^{\infty}|g^\d(r)|^2\, dr-\sin (m\phi)\int_0^{\infty}r^m|g(r)|^2\,dr\nonumber\\
&-&\int_0^{\infty}\left[\alpha r^2 \sin\left(3\phi\right)
+\beta r\sin\left(2\phi\right) +\gamma  \sin \phi\right] r |g(r)|^2\,dr\label{eq60}\\
&=&- \Im \left(\omega^2E_{*} \right)\int_0^{\infty}|g(r)|^2\, dr,\nonumber
\end{eqnarray}
where $\phi=\frac{2\pi}{m+2}-\theta.$ (See the proof of Theorem \ref{main} for the definition of $g(r)$.) 

Recall that the positivity of the left-hand side of (\ref{eq60}) implies $\lambda\in \R$. Since $|\theta|<\frac{\pi}{m+2}$, we get $\frac{\pi}{m+2}<\phi<\frac{3\pi}{m+2}$. Then since we are trying to show the left-hand side is positive under certain conditions on the coefficients, we restrict $\phi$ to $\frac{\pi}{m}\leq\phi\leq\frac{2\pi}{m}$ if $m\geq 5$, and $\frac{\pi}{4}\leq\phi<\frac{\pi}{2}$ if $m=4$, so that $\sin(m\phi)\leq 0$ in the second term above. (Note that when $m=4,\,\phi=\frac{\pi}{2}$, we have $\theta=-\frac{\pi}{6}$ for which some terms in (\ref{eq60}) are not integrable.)
We further want the discriminant of the quadratic  $\left[\alpha r^2 \sin\left(3\phi\right)
+\beta r\sin\left(2\phi\right) +\gamma  \sin \phi\right]$ to satisfy 
\begin{equation}\nonumber
\beta^2\sin^2(2\phi)-4\alpha\gamma \sin (3\phi)\sin\phi\leq 0,
\end{equation}
so that the quadratic expression has a single sign. That is, we want
\begin{equation}\label{quadratic}
\beta^2\leq 4\alpha\gamma \frac{\sin (3\phi)\sin\phi}{\sin^2(2\phi)}=\alpha\gamma\left(3-\tan^2\phi\right).
\end{equation}
 So  in order to have a positive right-hand side in (\ref{quadratic}), since $\alpha<0$ and $\gamma<0$, we need  $\phi\in \left[\left.\frac{\pi}{m},\,\frac{\pi}{3}\right)\right.\cap\left[\frac{\pi}{m},\,\frac{2\pi}{m}\right]$ for which   $\left(3-\tan^2\phi\right)$ is positive.
 Since $\left(3-\tan^2\phi\right)$ is decreasing, to maximize the right-hand side of (\ref{quadratic}), we choose $\phi=\frac{\pi}{m}$. Hence as we remarked at the beginning of the proof, this proves the reality of the eigenvalue under (\ref{eq51}).

Similarly, in order to prove the positivity of the eigenvalues, suppose $\lambda\in \R$ and use (\ref{sine-value}). Let $h(r)=v(re^{i\theta})=u(-ire^{i\theta})$.  Then  (\ref{sine-value}) becomes
\begin{eqnarray}
&&\sin\theta\int_0^{\infty}|h^\d|^2\,dr-\sin(m+1)\theta \int_0^{\infty}r^m|h|^2\,dr\nonumber\\
&-&\int_0^{\infty}\left[\alpha r^2 \sin\left(4\theta\right)
+\beta r\sin\left(3\theta\right) +\gamma  \sin \left(2\theta\right)\right] r |g(r)|^2\,dr\label{pos_prov}\\
&&=\lambda \sin \theta \int_0^{\infty}|h|^2\,dr,\quad\text{for all}\quad \frac{\pi}{m+2}<\theta<\frac{3\pi}{m+2}.\nonumber
\end{eqnarray}
Then we restrict $\theta\in \left[\frac{\pi}{m+1},\,\frac{2\pi}{m+1}\right]$ so that $\sin(m+1)\theta\leq 0$ in the second term above. We also want
 the discriminant of the quadratic
$\left[\alpha r^2 \sin\left(4\theta\right)
+\beta r\sin\left(3\theta\right) +\gamma  \sin \left(2\theta\right)\right]$ 
is nonnegative, so that  the quadratic expression has a single sign. That is, 
$$\beta^2\sin^2(3\theta)-4\alpha\gamma \sin (4\theta)\sin(2\theta)\sin^2(3\theta)\leq 0.$$  
Hence we have
\begin{equation}\label{disc}
\beta^2\leq 4\alpha\gamma\frac{ \sin (4\theta)\sin\left(2\theta\right)}{\sin^2(3\theta)}=32\alpha\gamma\frac{1-\tan^2\theta}{\left(3-\tan^2\theta\right)^2}.
\end{equation}
One can check $\frac{1-\tan^2\theta}{\left(3-\tan^2\theta\right)^2}$ is decreasing on $\left[\left.\frac{\pi}{m+1},\,\frac{\pi}{3}\right)\right.$. 
Also  we want to have $\left(1-\tan^2\theta\right)\geq 0$ so that the right-hand side of (\ref{disc}) is nonnegative (that is, $\frac{\pi}{m+1}\leq \theta\leq \frac{\pi}{4}$.) Then it is not difficult to see that $\theta=\frac{\pi}{m+1}$ maximizes the right-hand side of  (\ref{disc}). Also with  $\theta=\frac{\pi}{m+1}$, the left-hand side of (\ref{pos_prov}) is positive and hence $\lambda>0$. So with help of Theorem \ref{main}, we conclude that all real eigenvalues are positive under the hypothesis (\ref{eq52}).

Still we must show that eigenvalues are positive  under (\ref{eq51}). We will do this by showing that  the hypothesis (\ref{eq51}) implies (\ref{eq52}). That is, we will show
\begin{equation}\label{bigger}
 3-\tan^2\left(\frac{\pi}{m}\right)< 32\frac{1-\tan^2\left(\frac{\pi}{m+1}\right)}{\left(3-\tan^2\left(\frac{\pi}{m+1}\right) \right)^2},\qquad\text{for all $m\geq 4.$}
\end{equation}
Since $\frac{1-\tan^2\theta}{\left(3-\tan^2\theta\right)^2}$ is decreasing and positive for $\theta\in \left[\left.0,\,\frac{\pi}{4}\right)\right.$, the right-hand side of (\ref{bigger}) is an increasing function of $m\geq 5$ and hence greater than or equal to the value at $m=5$ which is $3$. So  (\ref{bigger}) holds for $m\geq 5$ since its left-hand side is less than $3$. And for $m=4$, one just check (\ref{bigger}) directly.
This completes the proof.
\end{proof}
\begin{remark}
{\rm Above we have chosen $P(z)=\alpha z^3+\beta z^2+\gamma z$ for simplicity. One should note that the above argument works for real polynomials of the type $P(z)=\alpha z^{n+k}+\beta z^{n}+\gamma z^{n-k}$ for some positive integers $n>k$.}
\end{remark}

The previous theorem handled $\alpha<0$. Similarly, we get the following for $\alpha> 0$ when $m=4,\,5,\,6$. 
\begin{theorem}\label{small_m}
Let $m=4,\,5,$ or $6$ and let $\alpha> 0,\, \gamma<0$. Suppose $\lambda$ is an eigenvalue of (\ref{ptsym}) with $P(z)=\alpha z^3+\beta z^2+\gamma z$ for some $\beta \in \R$. Then the eigenvalue is positive real, provided that 
\begin{equation}\label{first_one}
\beta\leq \sqrt{\alpha|\gamma|}\sqrt{\tan^2\left(\frac{2\pi}{m}\right)-3}.
\end{equation}
The eigenvalue $\lambda$ is also positive real provided that $\lambda\in \R$ and 
\begin{eqnarray}
\beta\leq  \left\{
                         \begin{array}{cl}

                        \infty & \quad \text{if $m=4,\,5,$}\\
                         &\\
                          4\sqrt{2}\sqrt{\alpha |\gamma|}\frac{\sqrt{1-\tan^2\left(\frac{2\pi}{7}\right)}}{3-\tan^2\left(\frac{2\pi}{7}\right)} & \quad \text{if $m=6$.}
                          \end{array}
                         \right. \label{second_one}
\end{eqnarray}
\end{theorem}
\begin{remark}
{\rm  In this theorem we restrict $m$ to  $m=4,\,5,\,6$ for reasons  explained in the proof below.

In (\ref{first_one}), by convention we take $\sqrt{|\alpha\gamma|}\sqrt{\tan^2\left(\frac{2\pi}{m}\right)-3}=+\infty$ when $m=4$, so that (\ref{first_one}) just says $\beta\in \R$, in that case, as in Theorem \ref{main}. Note that by (\ref{second_one}), all real eigenvalues are positive  when $m=5,\,\alpha>0,\,\beta\in \R,\,\gamma<0$. But there could perhaps be some non-real eigenvalues.

We mention that the case $\beta \leq 0$ or $m=4$ of Theorem \ref{small_m} is contained in Theorem \ref{main}, as it is explained in the proof of Theorem \ref{ext_thm}. The other cases of Theorem \ref{small_m} are new.}
\end{remark}
\begin{proof}[Proof of Theorem ~\ref{small_m}]
Since the case $\beta\leq 0$ is known already, it is enough to prove the theorem under the hypotheses (\ref{first_one}) and (\ref{second_one}) with $\beta$ replaced by $|\beta|$.

The proof below will be very much similar to that of the previous theorem. So we will refer equations to those in the proof of the previous theorem.
 
Since the case $m=6$ in (\ref{first_one}) says $\beta\leq 0$, this case is contained in Theorem \ref{main}. So for (\ref{first_one}) we can assume $m=5$.
Again we use (\ref{im_part}) with $P(z)=\alpha z^3+\beta z^2+\gamma z$. Then as we did in the proof of Theorem \ref{ext_thm}, we can get (\ref{eq60}), where we want  $\frac{\pi}{m}\leq\phi\leq\frac{2\pi}{m}$ so that $\sin(m\phi)\leq 0$ in the second term.  But this time since $\alpha>0$, we want $\sin (3\phi)<0$ and want the discriminant of the quadratic  $\left[\alpha r^2 \sin\left(3\phi\right)
+\beta r\sin\left(2\phi\right) +\gamma  \sin \phi\right]$ to be non-positive. Then again we have (\ref{quadratic}). Obviously we want a nonnegative right-hand side in (\ref{quadratic}), and  since $\alpha>0$ and $\gamma <0$, we need $\left(3-\tan^2\phi\right)\leq 0$. This means $\frac{\pi}{3}\leq\phi< \frac{\pi}{2}$ as well as $\frac{\pi}{m}\leq\phi\leq\frac{2\pi}{m}$.

Then in order to maximize the right-hand side of (\ref{quadratic}) we choose $\phi=\frac{2\pi}{5}$ for $m=5$. This proves the reality of the eigenvalue since $\beta\leq 0$ is covered by Theorem \ref{main}. 

Similarly, in order to prove the positivity of the eigenvalues under the hypothesis (\ref{second_one}), suppose $m=5$ or $6$, let $\lambda\in \R$ and use (\ref{sine-value}). Then like before we get (\ref{disc}) where we want  $\theta\in \left[\frac{\pi}{m+1},\,\frac{2\pi}{m+1}\right]$ so that $\sin(m+1)\theta\leq 0$ in the second term in  (\ref{pos_prov}). We further want the right-hand side of   (\ref{disc}) to be nonnegative, and hence want $\phi\geq \frac{\pi}{4}$. For $m=5$ since $\frac{\pi}{4}<\frac{\pi}{3}\leq \frac{2\pi}{m+1}$, we get $\beta< \infty$ from (\ref{disc}). And  for $m=6$ since $\frac{\pi}{4}<\frac{2\pi}{m+1}=\frac{2\pi}{7}<\frac{\pi}{3}$, and since $\frac{1-\tan^2\theta}{\left(3-\tan^2\theta\right)^2}<0$ is decreasing for $\theta\in \left[\left.\frac{\pi}{4},\,\frac{\pi}{3}\right)\right.$, we choose $\theta=\frac{2\pi}{7}$ in order to maximize the right-hand side of (\ref{disc}).
 Thus along with Theorem \ref{main} for $\beta\leq 0$, we conclude all real eigenvalues are positive under the hypotheses $\lambda\in \R$ and (\ref{second_one}).

Finally, it is not difficult to see that (\ref{first_one}) implies (\ref{second_one}), and hence we get the positivity of the eigenvalue under  (\ref{first_one}) as before. This completes the proof.
\end{proof}
The second method for sharpening Theorem \ref{main} is to make use of the $\int |g^\d(r)|^2\, dr$ term, by means of the harmonic oscillator inequality. 
Below we will prove that  eigenvalues $\lambda$ are positive real if  $\alpha < \frac{m}{2}$. Note that Dorey et al. \cite{Dorey} already prove this, and they also show that  eigenvalues are real if  $\alpha < \frac{m}{2}+2$.
\begin{theorem}\label{exactly_solv}
Let $m\geq 4$ be an even integer.  Suppose that an entire function $u$ along with $\lambda\in \C$ solves
\begin{equation}\label{eigen}
\left[-\frac{d^2}{dz^2}-(iz)^{m}-\alpha (iz)^{\frac{m}{2}-1}\right]u(z)=\lambda u(z),
\end{equation}
with the boundary conditions that $u$ decays to zero as $z$ tends to infinity along the rays $\arg z=-\frac{\pi}{2}\pm \frac{2\pi}{m+2}$. Then the eigenvalue $\lambda$ is  positive real if  $\alpha < \frac{m}{2}$. If $\alpha = \frac{m}{2}$, then all eigenvalues are positive real except the smallest one, which is zero and has the corresponding eigenfunction 
\begin{equation}\nonumber
u_0(z)=\exp\left[\frac{2}{m+2}(iz)^{\frac{m+2}{2}}\right].
\end{equation}
\end{theorem}
Note  that the function $v(z)=\exp\left[-\frac{2}{m+2}z^{\frac{m+2}{2}}\right]$ solves  $-v^\dd(z)+z^mv(z)=\frac{m}{2}z^{\frac{m}{2}-1}v(z)$ and decays along both ends of the real axis. This type of problems was studied by Bender and Wang \cite{Wang}.
\begin{proof}
The outline of the proof is similar to those of the proofs of Theorems \ref{ext_thm} and \ref{small_m} above. But this time we will make use of the $\int |g^\d|^2\, dr$ term via the harmonic oscillator inequality.

Like before, we examine the equation (\ref{im_part}), for the choice $P(iz)=\alpha (iz)^{\frac{m}{2}-1}.$ Then we have 
\begin{eqnarray}
&&\sin\left(\frac{4\pi}{m+2}-2\theta\right) \int_0^{\infty}|g^\d(r)|^2\, dr+\sin\left(m\theta+ \frac{4\pi}{m+2}\right)\int_0^{\infty}r^m|g(r)|^2\, dr\nonumber\\
&-&\alpha\sin\left(\frac{4\pi}{m+2}+\frac{m-2}{2}\theta\right)\int_0^{\infty} r^{\frac{m}{2}-1} |g(r)|^2\,dr\nonumber\\
&&\qquad\qquad=- \Im \left(\omega^2E_{*} \right)\int_0^{\infty}|g(r)|^2\, dr,\label{harmonic_ineq}
\end{eqnarray} 
where we refer to the proof of Theorem \ref{main} for the definition of $g(r)$.
Now we use the harmonic oscillator inequality on the first two terms above so that we have 
that for $|\theta|<\frac{\pi}{m+2}$ with $\sin\left(m\theta+ \frac{4\pi}{m+2}\right)\geq 0$,
\begin{eqnarray}
&&\sin\left(\frac{4\pi}{m+2}-2\theta\right) \int_0^{\infty}|g^\d(r)|^2\, dr+\sin\left(m\theta+ \frac{4\pi}{m+2}\right)\int_0^{\infty}r^m|g(r)|^2\, dr\nonumber\\
&\geq& \sqrt{\sin\left(\frac{4\pi}{m+2}-2\theta\right)\sin\left(m\theta+ \frac{4\pi}{m+2}\right)}\int_0^{\infty}2\,r^{\frac{m}{2}}|g^\d(r)g(r)|\, dr\nonumber\\
&\geq& \sqrt{\sin\left(\frac{4\pi}{m+2}-2\theta\right)\sin\left(m\theta+ \frac{4\pi}{m+2}\right)}\left|\int_0^{\infty}r^{\frac{m}{2}}\frac{d}{dr}|g(r)|^2\, dr\right|\nonumber\\
&=&\frac{m}{2}\sqrt{\sin\left(\frac{4\pi}{m+2}-2\theta\right)\sin\left(m\theta+ \frac{4\pi}{m+2}\right)}\int_0^{\infty}r^{\frac{m}{2}-1}|g(r)|^2\, dr,\nonumber
\end{eqnarray} 
 by parts. We then combine this with (\ref{harmonic_ineq}) to get that if
\begin{equation}\label{cond_sign}
\frac{m}{2}\sqrt{\sin\left(\frac{4\pi}{m+2}-2\theta\right)\sin\left(m\theta+ \frac{4\pi}{m+2}\right)}\geq\alpha\sin\left(\frac{4\pi}{m+2}+\frac{m-2}{2}\theta\right),
\end{equation}
then  $\Im \left(\omega^2E_{*} \right)\leq 0$, which then proves the reality of the eigenvalues like in the proof of Theorem \ref{main}.
Next we examine the condition (\ref{cond_sign}) and find with a little effort that $\theta=0$ is the best choice to get the best bound for $\alpha$ out of  (\ref{cond_sign}). That is, $\alpha\leq \frac{m}{2}$.

Similarly, in order to prove the positivity  and non-negativity of the eigenvalues, we use (\ref{sine-value}). Let $h(r)=u(-ire^{i\theta})$. Since  $\lambda\in \R$, one can get the following from (\ref{sine-value}):
\begin{eqnarray}
&&\sin\theta\int_0^{\infty}|h^\d|^2\,dr-\sin(m+1)\theta\int_0^{\infty}r^m|h(r)|^2\, dr-\alpha \sin\left(\frac{m}{2}\theta\right) \int_0^{\infty}r^{\frac{m}{2}-1}|h|^2\,dr\nonumber\\
&&=\lambda \sin \theta \int_0^{\infty}|h|^2\,dr,\qquad\quad\text{provided}\quad \frac{\pi}{m+2}<\theta<\frac{3\pi}{m+2}.\nonumber
\end{eqnarray}
Then since $\sin(m+1)\theta<0$ for $\frac{\pi}{m+1}<\theta<\frac{2\pi}{m+1}$, we apply the harmonic oscillator inequality to the first two terms above to get
\begin{eqnarray}
&&\lambda \sin \theta \int_0^{\infty}|h|^2\,dr\nonumber\\
&\geq& \left[\frac{m}{2}\sqrt{\sin\theta\left|\sin(m+1)\theta\right|}-\alpha \sin\frac{m}{2}\theta\right]\int_0^{\infty}r^{\frac{m}{2}-1}|h|^2\,dr\nonumber\\
&\geq& \left(\frac{m}{2}-\alpha\right) \int_0^{\infty}r^{\frac{m}{2}-1}|h|^2\,dr,\nonumber
\end{eqnarray}
where we have chosen $\theta=\frac{2\pi}{m+2}$. (One can check with a little effort that $\theta=\frac{2\pi}{m+2}$ is the best choice for this argument.) Since $\sin \theta >0$, we see $\lambda>0$ when $\alpha<\frac{m}{2}$ and $\lambda\geq 0$ when $\alpha=\frac{m}{2}$.

When $\alpha=\frac{m}{2}$, we see that for $m$ even,  $u_0(-iz)$ solves $-v^\dd(z)+\left(z^m+\frac{m}{2}z^{\frac{m}{2}-1}\right)v(z)=0$ with properties that  $u_0(-iz)$ decays in $S_{-1}\cup S_1$ and blows up in $S_0$. So it satisfies the proper boundary conditions of (\ref{rotated}) and hence $u_0(z)$ is, in fact, the eigenfunction of (\ref{eigen}).  Hence, all eigenvalues are positive except the smallest eigenvalue zero since eigenvalues are simple by Proposition \ref{main2}. This completes the proof.
\end{proof}

\begin{remark}
{\rm Note that arguments similar to Theorem \ref{exactly_solv} work, for example, for some polynomial potentials with $P(z)=\beta z^2+\gamma z$ when $m\geq 7$. Note also that to prove the reality and positivity of the eigenvalues, it is enough to show the left-hand sides of  (\ref{im_part}) and (\ref{sine-value}) are positive, respectively. For specific potentials, this might be achieved by other kinds of estimates.}
\end{remark}

\begin{remark}
{\rm The methods of proving the theorems in this section show how to sharpen Theorem \ref{main} to problems with potentials ``almost the same'' as those of Theorem \ref{main}. But for Theorem \ref{exactly_solv},  the proof of the reality can be shortened as follows.

Suppose that an analytic function $v(z)=f(z,G^{-1}(\alpha),E_*)$ along with $E_*\in \C$ 
solves 
$$-v^\dd(z)+(z^m-\alpha z^{\frac{m}{2}-1})v(z)=-E_*v(z), \quad \text{with}\quad v^\d(0)\overline{v(0)}=0,\quad v(+\infty+0i)=0.$$
This eigenproblem on the positive real axis is self-adjoint, and so $E_*\in \R$. Precisely,
 multiplying both sides by $\overline{v(z)}$, integrating over the positive real axis, and integrating the first term of the resulting equation by parts give 
$$\int_0^{\infty}|v^\d(x)|^2\,dx+\int_0^{\infty}x^m|v(x)|^2\,dx-\alpha \int_0^{\infty}x^{\frac{m}{2}-1}|v(x)|^2\,dx=-E_*\int_0^{\infty}|v(x)|^2\,dx.$$
(Hence $E_*\in \R$.) Then one uses the harmonic oscillator inequality on the first two terms to have 
$$ \left(\frac{m}{2}-\alpha\right)\int_0^{\infty}x^{\frac{m}{2}-1}|v(x)|^2\,dx\leq-E_* \int_0^{\infty}|v(x)|^2\,dx.$$
So if $\frac{m}{2}\geq\alpha$ then  $ E_*\leq 0$ which implies $\Im \left(\omega^2 E_*\right)\leq 0$. Thus the eigenvalue $\lambda$ is real, as before. }
\end{remark}
\section{Conclusions}
\label{conclusions}
In this paper we have proved that a family of one dimensional
 Schr\"odinger equations with $\mathcal{PT}$-symmetric polynomial potential $-[(iz)^m+a_1 (iz)^{m-1}+a_2 (iz)^{m-2}+\cdots+a_{m-1} (iz)]$  has all positive real eigenvalues, provided  that $(j-k)a_k\geq 0$ for all $k$, for some $1\leq j\leq \frac{m}{2}$. In particular, this result implies the original Bessis and Zinn-Justin conjecture for the potential $i\alpha z^3+z^2$.  

One would like  to further extend the proof of the reality and positivity to a larger class of  $\mathcal{PT}$-symmetric potentials. For example, can we get a similar conclusion for $j>\frac{m}{2}$? Also an interesting question will be how much the reality and positivity of the eigenvalues depend on the boundary conditions:  one certainly has some restrictions on choosing the boundary conditions (see, for example \cite{Bender2}). Our boundary conditions allow only one blowing up Stokes sector between the two decaying sectors near the negative imaginary axis. It will be also interesting to consider three or more (an odd number of) blowing up sectors between the two decaying sectors on which we impose the boundary conditions, with  the decaying sectors being symmetric with respect to the imaginary axis. Also the problem with the potentials $+(iz)^m-P(iz)$ (whose leading term has the opposite sign to those in the class of the problems studied in this paper) would be interesting too, in which case we impose the boundary conditions to allow an even number of blowing up sectors between the decaying sectors. Also it should be possible to apply the methods of this paper to some rational potentials, too.

One big question needing to be answered is to determine the span of the set of the eigenfunctions. For Sturm-Liouville problems, we know that zeros of eigenfunctions interlace, which seems to play an important role in the completeness of the set of the eigenfunctions.  Numerical work of Bender et al. \cite{Bender3} shows some intriguing interlacing properties of the zeros of the eigenfunctions for some $\mathcal{PT}$-symmetric oscillators, too. So understanding these interlacing properties of the zeros might lead us to progress. But yet, it seems there are a lot more questions than answers in this direction.

Also, one would like to study similar problems in higher dimensions.
\subsection*{{\bf Acknowledgments}}

The author was partially supported by the Campus Research Board at the University of Illinois. He thanks Richard S. Laugesen for encouragement, invaluable suggestions and discussions throughout the work.

{\sc email contact:}  kcshin@math.uiuc.edu
\end{document}